# Physically Constrained Ensemble Gaussian Process Modelling for Expensive Quantum Systems with Heteroskedastic Noise

Arpan Biswas[1,2], Sutirtha Paul[2,3], Joseph Agada[4], Matthias Thamm[5], Adrian Del Maestro[2,3]

[1]University of Tennessee-Oak Ridge Innovation Institute, University of Tennessee, Knoxville, TN 37996, USA

[2]Institute for Advanced Materials and Manufacturing, University of Tennessee, Knoxville, Tennessee 37996, USA

[3]Department of Physics and Astronomy, University of Tennessee, Knoxville, TN 37996, USA.

[4]Bredesen Center for Interdisciplinary Research, University of Tennessee, Knoxville, USA, 37996

[5]Institut für Theoretische Physik, Universität Leipzig, 04103 Leipzig, Germany

**Email**: abiswas5@utk.edu, Adrian.DelMaestro@utk.edu

*Accurate modeling of quantum many-body systems often requires computationally expensive simulations such as Density Matrix Renormalization Group (DMRG) or Quantum Monte Carlo (QMC) calculations. These methods, while precise, impose significant time and resource constraints, limiting their use in exhaustive parameter exploration. Moreover, these expensive simulations can contain variable errors over the large unknown parameter space, which needs to be quantified and propagated. Thus, predictive modelling is required to estimate the functional space accurately over scarcely sampled data with heteroskedastic noise, while preserving the physical relevance of the estimation. Therefore, we present a Physically Constrained Ensemble Gaussian Process (pc-EGP) framework designed to efficiently model complex and noisy quantum systems under physical consistency constraints. The proposed method first enforces physical constraints as a user controlled weighted penalty to the data-driven loss function of the Gaussian Process (GP) surrogates. Then an ensemble of such GP models is trained with variable noisy simulations via numerical quadrature method where these multiple GP(s) at different nodes is integrated as a quadrature weighted average. We first demonstrate the framework on synthetically generated data before applying to quantum systems. In the first case study, we leverage DMRG simulations of the Bose-Hubbard Model to predict the critical interaction parameter* $U_c$ *governing the superfluid-to-Mott-insulator transition. In the second case study, we demonstrate our method on QMC simulations, of a quantum liquid confined inside a nanoporous silicate with the goal of optimizing a chemical environment to realize a one-dimensional superfluid. Compared to conventional GP, pc-EGP achieves a better balance of accuracy and physically meaningful predictions. The proposed approach establishes a route towards robust and domain-informed autonomous and physically interpretable surrogate modeling of any computationally expensive system.*

**Keywords**: Gaussian Process, Bayesian optimization, physical constraint, heteroskedastic noise, quantum systems.

## 1. INTRODUCTION

The modern era of scientific discovery brings forth the challenge of exploring complex, large and time-consuming multidimensional domain specific parameter and function spaces. To accelerate the search for optimal parameter regimes, much effort has been focused on designing autonomous workflows via machine learning (ML)-driven selection (active learning). However, the current state of the art of surrogate modelling in such autonomous design is heavily assumed on the philosophy of continual learning from a large volume of data. These data-intensive approaches are not feasible in many real-world problems. As a motivating example, in simulations of microscopic models used to describe quantum systems of interacting bosons, due to an exponentially growing number of states, an exact computation of the superfluid properties are limited to $N \approx 20$[1] while much larger systems (N = 100-1000) are required to explore transitions between different quantum phases. These quantum phases can be defined as a function of several tuning parameters of the microscopic model. In order to explore over these tuning parameters to find an optimal point where a phase transition occurs, it may require millions of CPU core hours even to tune a single parameter (1D search space) using state-of-the art approaches, such as Density Matrix Renormalization Group (DMRG)[2–4] and quantum Monte Carlo (QMC)[5,6]. Moreover, for the surrogate modelling to learn the structure-property relationship of an expensive physical system (e.g. phase transitions), several physical constraints such as positivity, fixed density, or exact symmetries of the wavefunction need to be satisfied to provide a meaningful discovery. Furthermore, these expensive models contain variable errors over the large unknown parameter space, thus having heteroskedastic noise. Therefore, training a simple data-driven surrogate model with such small data volume will fail to accurately predict the quantum function spaces due to following simple prior assumptions of a continuous, noiseless or low homoskedastic or fixed noisy search space. This causes AI-misalignment with physical constraints. When ignored, it can lead to false predictions or unachievable discovery, ultimately de-accelerating (time and cost) the entire research process. In order to address such AI-misalignment in scientific problems where only small data is achievable, this paper aims to improve the gaussian process (GP) based surrogate modelling for heteroskedastic noisy and expensive data acquisition, with domain-specific physical constraint integration. Thus, we design a *physically-constrained ensemble Gaussian process model (pc-EGP)*.

The core of an autonomous workflow is a closed optimization or discovery loop, starting from performing an expensive experiment or simulation, measuring outcomes, parameter-outcome relationship analysis by AI/ML and decision-making algorithms to select the next experiment/simulation. One of the most popular approaches in the design of Autonomous Experiments (AE) is Bayesian optimization (BO). BO[7–9] is an active learning method which aims to autonomously explore the parameter space and continually learns the unknown ground truth and its global optimal region, where the ground truth function can be either black-box or expensive to evaluate, or both. Given a few evaluated training samples, the expensive unknown function is replaced by a cheaper surrogate model (*e.g.* Gaussian Process)[10–12], and the surrogate model continues to learn the region of interests with adaptive selection (*e.g.* Acquisition function)[13–16] of locations for future expensive evaluations. BO is more popular than other designs of experiment methods (*e.g.* Random sampling, Latin Hypercube sampling, *etc.*) as it is designed to converge to optimality with minimal number of expensive evaluations. Thus, BO has attracted special attention in the materials science domain where accelerated BO driven autonomous discovery has been particularly impactful, enabling efficient identification of optimal conditions for particular material properties without human intervention, such as bandgap optimization[17], small-molecule emitters discovery, maximizing carbon nanotube growth rates[18], and so on. This type of autonomous workflow with BO has been widely used in recent studies to adaptively explore expensive control parameter spaces of physical/simulation models[19–24] and experiments[25], material structure-property relationship discovery[26], and to develop autonomous platforms towards accelerating chemical[27,28] and material design[29–31]. A number of excellent reviews on BO are available[7,32], and it is now implemented in a broad range of Python libraries including BOtorch (Pytorch)[33] and Gpax (Jax)[34].

One of the key elements of the success of BO is the surrogate modelling, where a Gaussian process is a popular method as it can capture the mean and the uncertainty of estimation over the unknown function space. Being an error-dependent kriging model, it provides higher confidence in estimation where explored training samples are denser, which helps to guide the BO efficiently where next sampling should be targeted for discovery. However, the standard GP based surrogate model is still purely data-driven, with no quality control features to guarantee the predictions are physically meaningful. Also, the standard GP often assumes fixed or homoskedastic noise in the expensive system. Thus, failure to provide physically meaningful predictions or capturing the true heteroskedastic noise efficiently, affect the future decisions in BO exploration. In this paper, we aim to improve the GP modeling architecture with 1) the integration of improved physical constraints and 2) an approach to propagate and quantify the heteroskedastic noise of the expensive physical systems into the surrogate estimation. We have demonstrated the proposed pc-EGP to (i) precisely identify the location of a quantum phase transition between a superfluid and insulator in the Bose-Hubbard model, and

(ii) to learn the chemical potential dependence of superfluid filling inside nanoporous materials.

The roadmap for this paper is as follows: Section 2 provides the detailed mathematical description and algorithm of the pc-EGP architecture. Section 3 showcases the implementation of the prototype to synthetic data, before applying to real QMC and DMRG simulation of low dimensional superfluids. Section 4 concludes the paper with final thoughts.

## 2. METHODOLOGY

The proposed pc-EGP model has two key developments- 1) the injection of physical constraints during the training process by jointly optimizing the model hyperparameters via a user-controlled integrated physical loss function, and 2) an ensemble modelling approach via numerical quadrature method to propagate and quantify heteroskedastic noise present in physical systems.

### *2.1. Physical Loss Integration into a Standard Gaussian Process (GP) model*:

The general form of the GP model is as follows:

$$y(x) = \Delta(x) + \varepsilon \qquad (1)$$

where $\varepsilon \sim N(0, \sigma_e^2 I)$ is the standard fixed noise model with zero mean and variance as the mean variance of the physical system. $\Delta(x)$ is a realization of a correlated Gaussian Process with mean $E[\Delta(x)]$ and covariance $k(x^i, x^j)$ kernel functions defined as follows:

$$\Delta(x) \sim GP\left(E[z(x)], k(x^i, x^j)\right); \quad (2)$$

$$E[\Delta(x)] = 0, k(x^i, x^j) = \sigma^2 R(x^i, x^j) \quad (3)$$

$$R(x^i, x^j) = k_\vartheta \times \left(\left\|\frac{x^i - x^j}{\theta}\right\|\right) \quad (4)$$

where $\sigma^2$ is the overall variance parameter and $\theta$ is the correlation length scale parameter. These are termed as the hyper-parameters of GP model. $R(x^i, x^j)$ is the spatial correlation function. In this paper, we have considered a Matern Kernel which is a generalization of the standard exponential kernel in **eq. 4**. The modified kernel functions at different smoothness parameter $\vartheta = 0.5, 1.5, 2.5$ are given as per **eqs. 5-7**.

$$R(x^i, x^j)_{\vartheta=0.5} = \exp\left(-\left\|\frac{x^i - x^j}{\theta}\right\|\right) \quad (5)$$

$$R(x^i, x^j)_{\vartheta=1.5} = (1 + \sqrt{3} \times \left\|\frac{x^i - x^j}{\theta}\right\|) \times \exp\left(-\sqrt{3} \times \left\|\frac{x^i - x^j}{\theta}\right\|\right) \quad (6)$$

$$R(x^i, x^j)_{\vartheta=2.5} = (1 + \sqrt{5} \times \left\|\frac{x^i - x^j}{\theta}\right\| + \frac{5}{3} \times \left(\left\|\frac{x^i - x^j}{\theta}\right\|\right)^2) \times \exp\left(-\sqrt{5} \times \left\|\frac{x^i - x^j}{\theta}\right\|\right) \quad (7)$$

It is to be noted that the integration of the physical constraint can be applicable to any choice of kernel function. The objective is to estimate (via Maximum Likelihood Estimation), the hyper-parameters $\sigma, \sigma_e^2, \theta$ which create the surrogate model that best explains the training data $\boldsymbol{D_n}$ at iteration $n$. Here, we integrate a physical loss component with the traditional data-driven loss function, which we aim to minimize. The physics driven negative log-likelihood loss function $L_{phy}$ can be stated as below:

$$L_{phy} = \frac{1}{2} y^T K^{-1} y + \frac{1}{2} \log|K| + \frac{N}{2} \log 2\pi \quad (8)$$

$$K = K_{XX} + \sigma_e^2 I + pI \quad (9)$$

$$p = w_1 l_1 + w_2 l_2 + \ldots + w_q l_q \quad (10)$$

where $K$ is the covariance matrix of the observation, $N$ is the total number of training samples, $p$ is the physical loss value, $[l_1, l_2, \ldots, l_q]$ are the $q$ loss component values controlled with trade-offs weights $[w_1, w_2, \ldots, w_q]$, $K_{XX} = k(\boldsymbol{X}, \boldsymbol{X})$ is the covariance matrix for training inputs, $I$ is the identity matrix of dimension $N$.

After the GP model is fitted, the next task of the GP model is to predict at an arbitrary (unexplored) location drawn from the parameter space. Assume $\boldsymbol{D_n} = \{\boldsymbol{X_n}, \boldsymbol{Y(X_n)}\}$ is the prior information from previous evaluations or experiments from high fidelity models, and $\bar{\bar{x}}_{n+1} \in \bar{\bar{\boldsymbol{X}}}$ is a new parameter choice within the unexplored locations in the parameter space, $\bar{\bar{\boldsymbol{X}}}$. The predictive output distribution of $x_{n+1}$, given the posterior GP model, is given by **eqs. 11-13.**

$$P(\bar{\bar{y}}_{n+1} | \boldsymbol{D_n}, \bar{\bar{x}}_{n+1}, \sigma, \sigma_e^2, \boldsymbol{\theta}) = \boldsymbol{N}(\mu(\bar{\bar{y}}_{n+1}(\bar{\bar{x}}_{n+1})), \sigma^2(\bar{\bar{y}}_{n+1}(\bar{\bar{x}}_{n+1}))) \quad (11)$$

where:

$$\mu(\bar{\bar{y}}_{n+1}(\bar{\bar{x}}_{n+1})) = \boldsymbol{K_{n+1}^T K_n^{-1} Y_n}; \quad (12)$$

$$\sigma^2(\bar{\bar{y}}_{n+1}(\bar{\bar{x}}_{n+1})) = diag\left(k(\bar{\bar{x}}_{n+1}, \bar{\bar{x}}_{n+1}) - \boldsymbol{K_{n+1}^T K_k^{-1} K_{n+1}}\right) \quad (13)$$

$\boldsymbol{K_n}$ is the kernel matrix of already sampled designs $\boldsymbol{X_n}$ and $\boldsymbol{K_{n+1}}$ is the covariance function of new design $\bar{\bar{x}}_{n+1}$ which is defined as follows:

$$\boldsymbol{K_n} = \begin{bmatrix} cov(x_1, x_1) & \cdots & cov(x_1, x_n) \\ \vdots & \ddots & \vdots \\ cov(x_n, x_1) & \cdots & cov(x_n, x_n) \end{bmatrix}$$

$$\boldsymbol{K_{n+1}} = [cov(\bar{\bar{x}}_{n+1}, x_1), cov(\bar{\bar{x}}_{n+1}, x_2), \ldots, cov(\bar{\bar{x}}_{n+1}, x_n)]$$

In standard GP, the prediction is not bound by non-negative values which can be infeasible to model various physical quantities including energy gaps or particle densities in quantum systems.

Thus, expanding **eq. 10**, we calculate the physical loss $l_1$ as per below **eq. 14.**

$$l_1 = \frac{1}{N_v + 1} \sum_{i \notin v} |\mu_i| \quad (14)$$

where $\nu$ is the probabilistically non-violated samples as validated from $\{i: P(\mu_i \geq 0) > R\}$ with Reliability Index $R = 0.95$, $N_\nu$ is the number of violating samples. It is to be noted, any form of other physical constraint can be implemented as another component of loss. Then, to ensure the hyperparameter still optimizes while trying to best fit with the training samples, a second loss function $l_2$ is required. We calculate the loss $l_2$, to minimize the difference between prediction and actual $n$ training samples, as per below **eq. 15**.

$$l_2 = \frac{1}{n}\sum_{i=1}^{n} |y_i - \mu_i| \quad (15)$$

where $y_i$ and $\mu_i$ are normalized with providing a slack value $\xi = 10^{-12}$ in the denominator for numerical stability.

### *2.2. Quantification of heteroskedastic noise via ensemble Gaussian Process*:

In this section, we explain the development of the ensemble GP model to quantify heteroskedastic noise from the physical system, via numerical quadrature method. The predictive means and variances of each unexplored sample, as per **eqs. 11-13**, only accounts for the GP model uncertainty, where the outcome $y$ of the training data $\boldsymbol{D_n}$ is considered deterministic (no variance). The uncertainty of the outcomes is integrated as mean variance into the fixed noise model $\varepsilon$. In this paper, we aim to design a noise model to consider the propagation of the noise, at each input point $x$. Revisiting the training samples outcomes with mean and variance gathered from the physical system (assuming gaussian distribution), we employ Gauss–Hermite quadrature as below **eq. 16.**

$$\int_{-\infty}^{\infty} f(z)e^{-z^2}dz \approx \sum_{j=1}^{m} \omega_j f(z_j) \quad (16)$$

where $z_j$ is the Hermite quadrature nodes, $\omega_j$ is the quadrature weights, $m$ is the number of nodes. With approximation from Gauss–Hermite quadrature, for each training sample $i$, the collocation nodes and corresponding weights are computed as per **eqs. 17-18.** Thus, each uncertain training sample outcome $y$ is represented by $m$ deterministic realizations without requiring expensive Monte Carlo sampling.

$$y_{i,j} = y_i + \sqrt{2}\sigma_{e,i} z_j \quad (17)$$

$$\omega_j = \frac{\omega_j}{\sqrt{\pi}} \quad (18)$$

Then, each deterministic realized training data, $\boldsymbol{D_{n,j}} = \{\boldsymbol{X_n}, \boldsymbol{Y_j}\}$ is fitted to earlier stated physically constrained GP model $\Delta_{\boldsymbol{j}}$, and the predictive mean and variance of the new unexplored point $\bar{\bar{x}}_{n+1}$ is calculated (following **eqs. 2-15**). Finally, the weighted predictive mean and the variance of the new unexplored point $\bar{\bar{x}}_{n+1}$ is calculated as below **eq. 19-20.** To avoid overestimation, we set $\sigma_e^2 = 0$ in the training module of $\Delta_{\boldsymbol{j}}$.

$$\mu(\bar{\bar{x}}_{n+1}) = \sum_{j=1}^{m} \omega_j \mu_j(\bar{\bar{x}}_{n+1}) \quad (19)$$

$$\sigma^2(\bar{\bar{x}}_{n+1}) = \sum_{j=1}^{m} \omega_j \sigma_j^2(\bar{\bar{x}}_{n+1}) \quad (20)$$

To demonstrate the proposed pc-EGP, we first tested over synthetically generated data. The first ground truth function is considered deterministic, where we compare the results between standard GP and physically constrained GP. During the training process, we considered the Adam optimizer with learning rate of 0.2 for the hyperparameter optimization to minimize the negative likelihood loss with total epochs of 150. We consider the Matern Kernel **eq. 6.** **Fig. 1** shows the performance of pc-GP. **Fig. 1a** is the deterministic synthetic function where the red dots are the initial training samples. For better sampling, we choose the next 30 samples autonomously via GP-BO and pc-GP-BO, both minimizing and maximizing the deterministic function to exploit more samples over low and high value regions respectively. The standard BO algorithm is provided in the Supplementary Material **Table A1**. **Figs. 1b, 1c** are the performance of the GP and pc-GP respectively as more sampling is completed over low function value region. It is to be noted, to add complexity, we intentionally choose the synthetic function where the low-valued region of interest has negative values. Like in many physical systems where the outcome over a sub-search space can be highly misaligned and we can obtain erroneous data due to stochastic sampling errors (e.g. QMC), our objective is to validate where the pc-GP will be able to provide improved predictions with combined with domain-expert driven physical constraints. As expected from **fig. 1b**, the standard GP always focuses on minimizing the error between prediction and actual value of the training samples. However, it fails to avoid erroneous samples (green dots under the red solid $y = 0$ horizontal line) and thus provide an erroneous prediction (blue solid line under the $y = 0$ horizontal line). The predicted minimum found is $\mu_{min} = -0.0848$. Looking into the performance of pc-GP, after tuning the trade-offs weights $[w_1 = 20, w_2 = 1]$, we see the prediction is aligned to the constraint, even when there are several erroneous training samples. Though the tradeoffs were majorly aligned to abide by the constraint, the pc-GP still ties with fitting the training data to best match with the prediction. Thus, we see the prediction line near the (BO-exploited) low function value region just above the constraint boundary without too much deviation from the training data. The predicted minimum found in this case is $\mu_{min} = 0.0155$. To summarize, we can see the power of integration of the domain-informed constraints to avoid non-physical estimation, even if the actual sampled is incorrect due to severe noise in the physical systems.

**Figs. 1d-1f** are the performance of the GP and pc-GP (without and with tuning trade-off weights) respectively as more sampling is done over the high function value region. It is to be noted, in any of the cases, we have not obtained

any erroneous samples (green dots under the red solid $y = 0$ horizontal line), as the region of interest is the maxima region. However, in most cases, one wants to learn the physical system as accurately as possible for the non-interesting region as well, without high sampling costs. We see that in all cases as predicted by GP and pc-GP, the respective predicted means (solid blue line) are always in the feasible region. This is expected as there are no erroneous training samples. However, as we can see from **fig. 1d**, the standard GP prediction has the largest deviations in the minima region as $\mu_{min} = 0.4911$. From **fig. 1e**, when the pc-GP is not tuned with trade-offs weights $[w_1 = 1, w_2 = 1]$, we see a slight improvement in prediction with $\mu_{min} = 0.46$. However, as we tuned the trade-offs weights $[w_1 = 1, w_2 = 3]$, we could see a significant improvement in the low-sampled region as well with $\mu_{min} = 0.1438$. This shows the significance of the constraint loss function $l_2$, in addition to the physical loss function such as $l_1$.

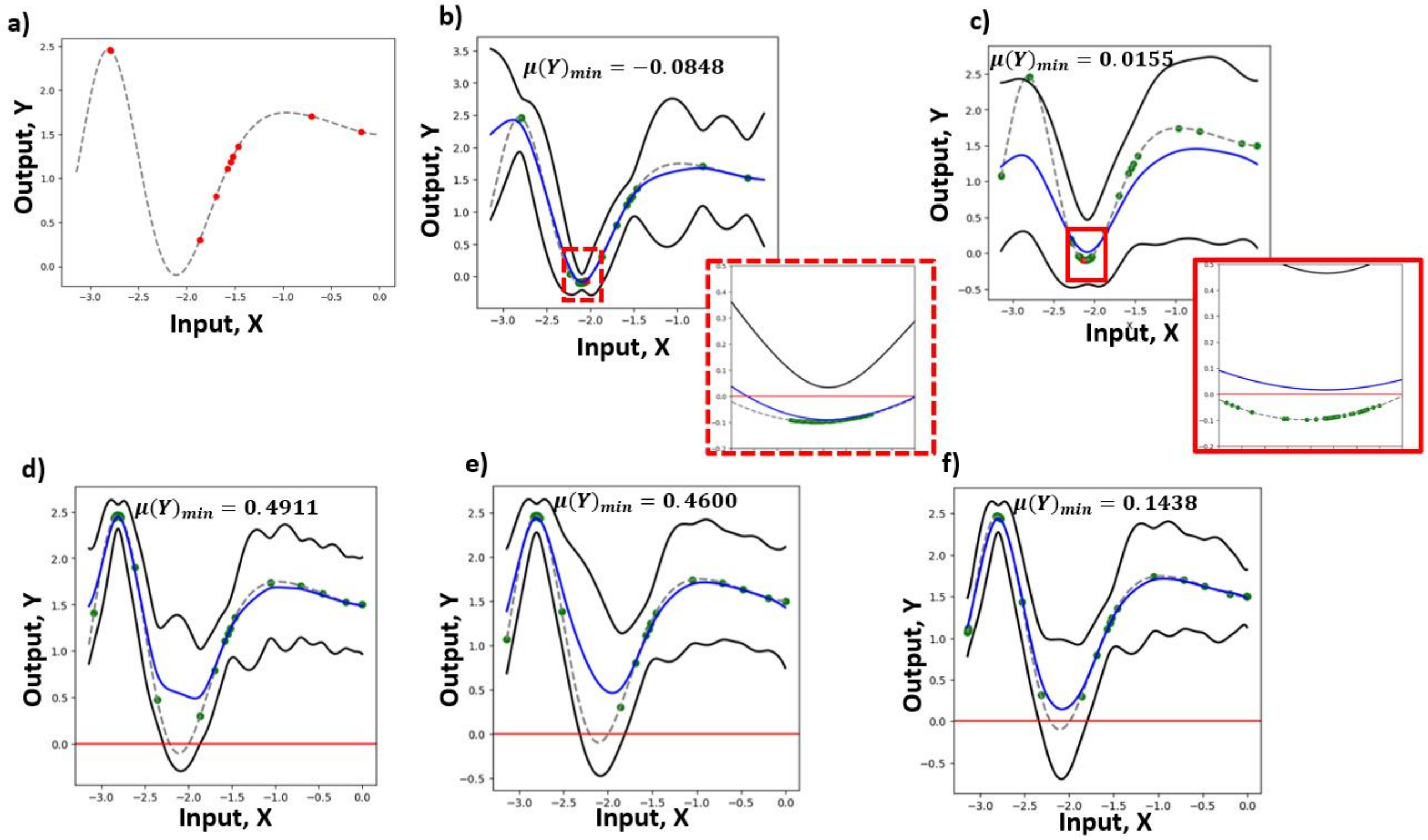

**Fig. 1.** Performance of pc-GP over synthetic test function. Fig 1(a) shows the deterministic test function where red dots are an initial 10 samples. Figs 1(b) (GP) and 1(c) (pc-GP) are the prediction (blue solid lines) from GP and pc-GP with tuned weights $[w_1 = 20, w_2 = 1]$ where an additional 30 samples have been generated via ***minimizing*** the test function with Bayesian optimization (BO). Fig 1(d) - 1(f) are the prediction (blue solid lines) from GP, pc-GP with untuned weights $[w_1 = 1, w_2 = 1]$ and tuned weights $[w_1 = 1, w_2 = 3]$ respectively where an additional 30 samples have been generated via ***maximizing*** the test function with BO. The dashed line is the test function, and all the training samples are highlighted as green dots. The black solid lines are the $\pm 2$ GP standard deviation. The red horizontal solid line is the applied constraint border line of $y \geq 0$.

### *2.3. Demonstration of pc-EGP over numerical test functions:*

To demonstrate the full potential of pc-EGP, we modified the test function with heteroskedastic noise as per in Supplementary material **Fig. A1.** Here, we considered $m = 5$ deterministic realizations, derived from the mean and variance of the test function. During the training process of each GP, we considered the Adam optimizer with learning rate of 0.2 for hyperparameter optimization with total epochs of 150. It is to be noted that this optimizer settings can be different and tuned independently. In the first case, we randomly generated 40 training samples as denoted by the red dots in **Figs 2a-2d.** Here, we have very few samples near the constraint boundary region (low-function value region). Though in each GP architecture, the predictions are always in the feasible region, the standard GP (**fig. 2a**) provides the worst prediction as farthest away from the data with $\mu_{min} = 0.2209$ . To reiterate, while one objective is not to entirely focus on fitting the data to predict in the infeasible region, another objective is to still predict close to that erroneous data while being in the feasible regime. Thus, with pc-GP (fig. 2b), we can achieve that with better prediction while in the feasible regime with $\mu_{min} = 0.1608$ with tuned weights of $[w_1 = 11, w_2 = 1]$. Next, with EGP (without constraints as in fig. 2c), we can see significant improvement in prediction near the constraint boundary region with $\mu_{min} = 0.0612$. This shows the significance of ensemble GP for better estimation than the standard GP, particularly for the region with less data. Finally, integrating the constraints with the EGP, the pc-EGP (fig. 2d) shows the best prediction with closest alignment with the training data, but within the feasible region as $\mu_{min} = 0.0059$.

In the second case, we randomly generated 100 training samples as denoted by the red dots in **Figs 2e-2h.** Here, we have high number of erroneous samples near the constraint boundary region (low-function-value region). Thus, we now see how the standard GP (**fig. 2e**) can provide infeasible prediction with $\mu_{min} = -0.0625$. Here the significance of the pc-GP comes into play (fig. 2f), as we can still achieve the prediction in the feasible regime with $\mu_{min} = 0.1186$ with tuned weights of $[w_1 = 35, w_2 = 1]$. Interestingly, now with EGP (without constraints as in **fig. 2g**), we can this the worst performance with providing higher infeasible predictions with $\mu_{min} = -0.1021$. However, this makes sense, as the purpose of the ensemble GP is to better fit the data than the standard GP, however, it is unknown to the feasibility of the training data. Thus, with pc-EGP (fig. 2h), we see the best prediction with closest alignment towards the training data, but within the feasible region as $\mu_{min} = 0.0003$, even though there are larger number of erroneous training data.

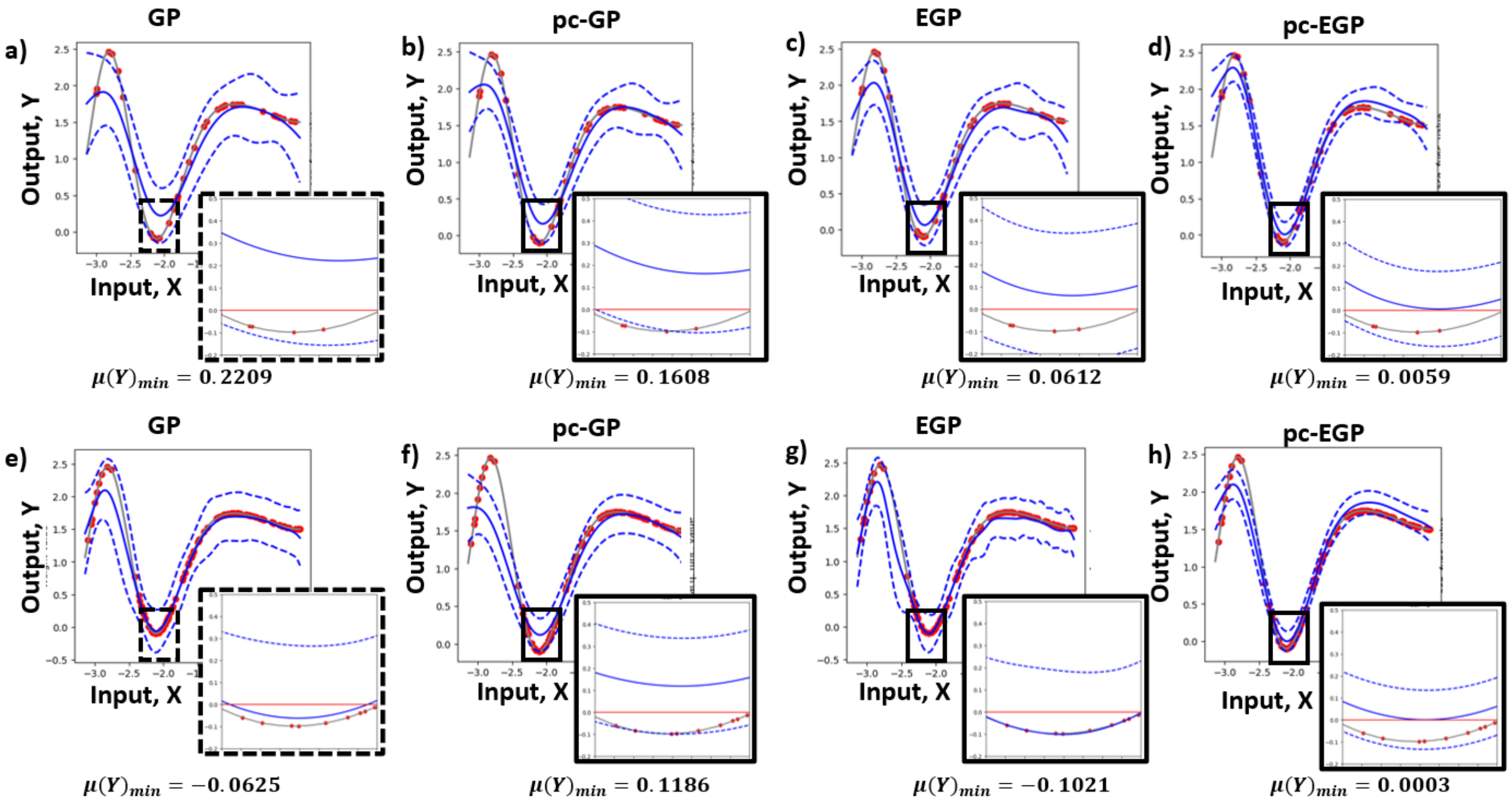

**Fig. 2.** Performance of pc-GP, EGP and pc-EGP for a synthetic test function with heteroskedastic noise. The test mean function is the same as fig 1(a) where the details on the expansion to the noisy function is provided in Supplementary Material Fig. A1. Figs 2(a) - 1(d) are the prediction (blue solid lines) from GP, pc-GP, EGP and pc-EGP respectively with 40 randomly selected training samples. For pc-EGP and pc-GP, the tuned weights are $[w_1 = 11, w_2 = 1]$. Figs 2(e) - 2(h) are the prediction (blue solid lines) from GP, pc-GP, EGP and pc-EGP respectively with 100 randomly selected training samples (red dots). For pc-EGP and pc-GP, the tuned weights are $[w_1 = 35, w_2 = 1]$. The solid gray line is the test mean function, and all the training samples are highlighted as red dots. The blue dashed lines are the $\pm 1$ GP standard deviation. The red horizontal solid line is the applied constraint $y = 0$.

## 3. RESULTS

From the above analysis over the synthetically generated functions, we can see promising results where the proposed method has the ability 1) to rectify the estimation from the enormous sampled data, 2) gaining higher confidence in estimation from the data with heteroskedastic noise and 3) balancing the physical and data-driven constraints for an improved and meaningful estimation. In support to these results, in this section, we demonstrate the application and performance of the proposed pc-EGP model over exploration and learning to two complex and expensive quantum systems which have been previously studied in the physics literature. The first is the precise numerical identification of the one-dimensional Bose-Hubbard model parameters that yield a quantum (non-thermal) phase transition[4], while the second involves a potential physical realization of this behavior where helium atoms are confined inside an ordered nanoporous glass[35–37].

**Fig 3.** shows the performance of the estimation of the proposed model pc-EGP, in learning the Bose–Hubbard model, leveraging DMRG and QMC simulations, to predict the critical interaction parameter $U_c/J$ governing the superfluid-to-Mott-insulator transition where $\left(1-\zeta(\frac{U}{J})\right)^2=0$. For this dataset, the noise heteroskedastic as each simulated value of $U/J$ has a different propagated uncertainty. Moreover, while the simulation tuning parameter is $U/J$, the region of interest and critical value $U_c/J$ depends on the noisy measurement of a proxy $\zeta$ whose functional dependence on $U/J$ is unknown. Thus, the goal is to learn where $(1-\zeta)^2=0$ from sparse data. This process is shown in **fig. 3a**, where we only have 9 DMRG data points (due to their extreme computational cost) with the region of interest near $(1-\zeta)^2=0$ is shown by the black box. However, as per physical and functional knowledge $(1-\zeta)^2\geq 0$ at any value of $U/J$ (see bottom panel of fig. 3a). During the training process of each GP, we considered Adam optimizer with learning rate of 0.2 for the hyperparameter optimization with total epoch of 150. We consider the Matern Kernel **eq. 6.** Estimating from the standard GP which was applied in the authors original study[4], as per **fig. 3b**, we can see that the mean function does not entirely goes through the DMRG simulated samples due to associated simulation uncertainty. However, the stated physical knowledge is violated where negative mean values are predicted in the region of interest, yielding an unphysical mean ${(1-\zeta)_{min}}^2=-1.2\times10^{-8}$ with uncertainty extending well into the forbidden region. On the other hand, estimating from pc-EGP with tuned weights of $[w_1=14, w_2=1]$ (**fig. 3c**), we observe no such violations where we find the mean ${(1-\zeta)_{min}}^2=4\times10^{-9}$. Interestingly, however, we see the variance is higher in pc-EGP estimation than in standard GP. This could be because the GP tries to only fit closer to the training data points, and thus the model feels more confidence in estimation. However, in the real case, this is not true as doing so causes physical constraint violations. In pc-EGP, the algorithm doesn't purely focus on minimizing the difference between predicted and training data points but co-minimizes user-defined physical violations. However, since the model variance focuses on how well the data fits to the training points and not on physically meaningful estimation, we can get sub-informative variances. Comparing the critical phase transition value $U_c/J$ over parameter space, the standard GP and pc-EGP estimate $\frac{U_c}{J}=3.2733$ and $\frac{U_c}{J}=3.2730$ respectively in close agreement within the uncertainty of the reported predictions of Ref[4].

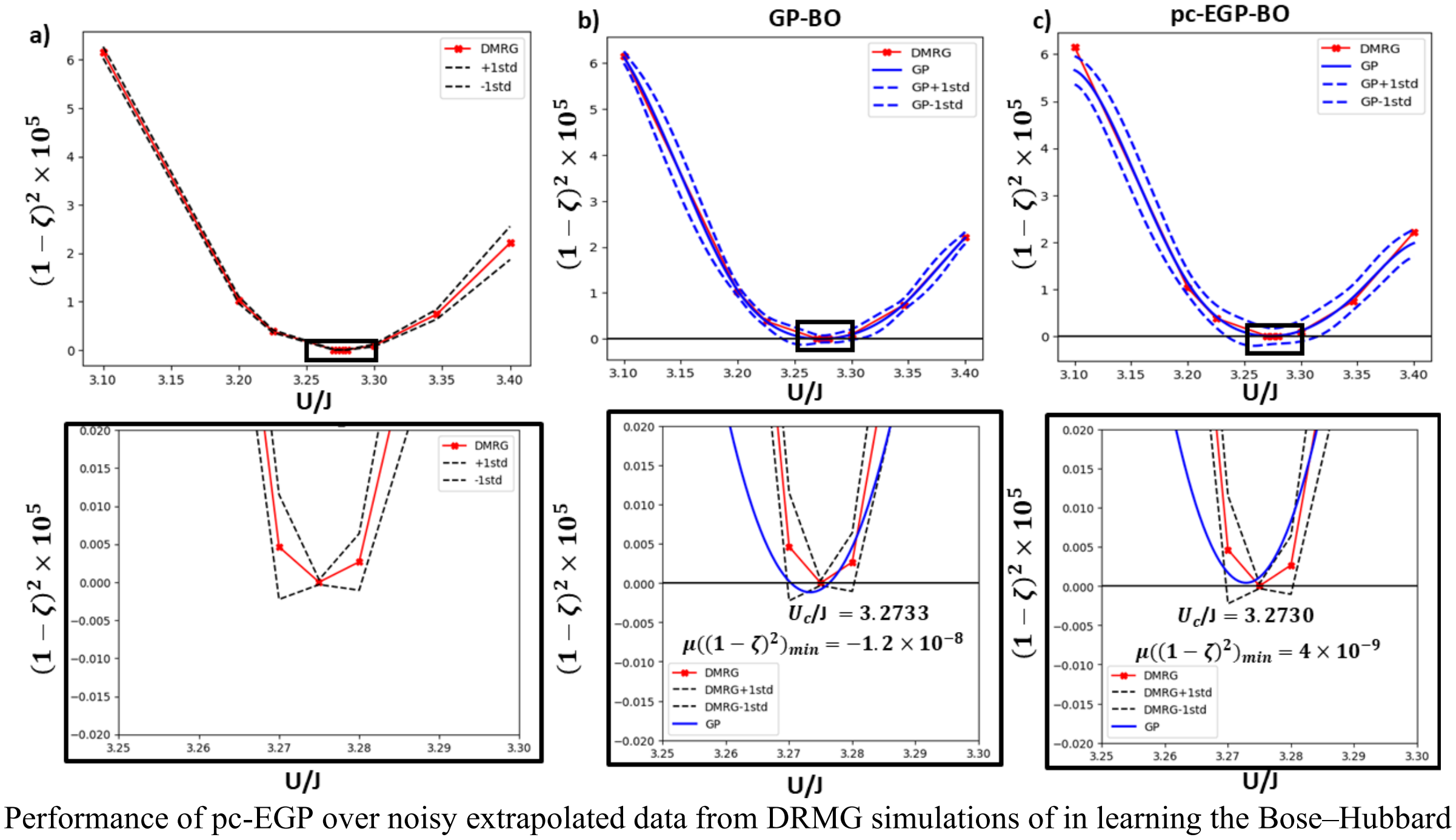

**Fig. 3.** Performance of pc-EGP over noisy extrapolated data from DRMG simulations of in learning the Bose–Hubbard model to predict the critical interaction parameter $U_c/J$ governing the superfluid-to-Mott-insulator transition. Fig 3(a) shows the expensive 9 DMRG samples in red line with dots. The black dashed lines are the $\pm1$ standard deviation of the DMRG simulations. Fig 3(b)-3(c) are the prediction (solid blue lines) and $\pm1$ standard deviation (dashed blue lines) from GP and pc-EGP respectively after fitting these 9 samples. For pc-EGP, the tuned weights are $[w_1=14, w_2=1]$.

As per our second case study, one possible physical realization of the one-dimensional Bose-Hubbard model would be to confine bosonic helium atoms ($^4$He) inside a nanoporous framework that yields confinement to quasi-one-dimension[35]. The challenge is that validating this prediction requires expensive QMC simulations to identify the experimental conditions (e.g. temperature, density, chemical potential/pressure, pore radius, pre-plating materials) where this behavior may manifest. A sparse subset of the total number of required QMC simulations for helium confined inside cylindrical nanopores was performed (details in Ref.[35]]) with the main observable being the radial dependence of the $^4$He density ($\rho$) inside the pore as a function of chemical potential. The goal is to learn the temperature ($T$), chemical potential ($\upsilon$) and external pore-radius ($R$) dependence of the radial density from sparse simulation data at these fixed parameters. We begin with a benchmark case where the ground truth value of this function $\rho(r; \upsilon, T, R)$ can be exactly determined at zero temperature ($T = 0$) and we try to learn its $r$ dependence from QMC simulation data with heteroskedastic noise. Here, the physical constraint is that the radial density must be non-negative for all parameter values $\rho \geq 0$.

**Fig 4.** shows the performance of autonomous exploration via the proposed model pc-EGP-BO. During the training process of each GP, we considered Adam optimizer with learning rate of 0.2 for the hyperparameter optimization with total epochs of 150. We consider the Matern Kernel **eq. 6. Fig 4a** shows the expensive sparsely sampled ground truth and noisy densely sampled QMC simulations as denoted by the red and gray lines. Here each individual simulation can take up to 10,000 core hours. The gray dashed lines are the $\pm 2$ standard deviation of noisy QMC simulations. Looking into the QMC data, we can see the noise is heteroskedastic with higher variance at high density region, and deterministic (no noise) in the low-density $\rho = 0$ region between $r = 5$ and $r = 8$. From domain-knowledge, we know that at any value of $r, \rho \geq 0$. It is to be noted, for the validation in having sufficient data for BO exploration, we aim to collect training data from noisy QMC simulations. Here, we considered $m = 5$ deterministic realizations, derived from the mean and error of the QMC simulations. During the training process of each GP, we considered the Adam optimizer with learning rate of 0.2 for hyperparameter optimization with total epochs of 150. We consider the Matern Kernel **eq. 6.** We started the BO with 10 randomly selected QMC simulated data as denoted by green dots and aim to explore the region over the parameter space of radius $r$ within the nanopore, having the highest value of density $\rho$ as this is where the simulations predict the most likely location for localized $^4$He atoms.

**Fig. 4b** shows the exploration of the traditional GP-BO after 30 iterations. We can see that even though we have autonomously sampled the high-density region, the prediction (solid blue line) is still significantly different than the QMC simulation (solid gray line). Additionally, the GP variance (area within dashed blue lines) is large due to inaccurate predictions, which show the model has low confidence in its prediction. Moreover, for large values of $r$ where it is expected that the atom density should be $\rho = 0$ due to energetic penalty, the model provides an unphysical prediction. Therefore, the absolute difference of the predicted highest and the lowest density from the actual simulated mean value are **0.1094** and **0.6079** respectively. **Fig. 4c** shows the exploration of the proposed pc-EGP-BO after 30 iterations, starting with the same initial samples and objective to exploit the high-density region over the parameter space. We can clearly see a significant improvement in the prediction in the high-density region with the model variance (area within dashed blue lines) staying within the variance of the QMC simulations (area within dashed gray lines). This suggests the pc-EGP provide predictions with higher confidence at the same cost of exploration. Similarly, the low-sampled $\rho = 0$ region is highly accurate with significantly low model variance. Though we find a very small violation with predicted minimum density $\mu_{min} = -0.0012$, this could be due to the optimizer not finding a better solution to balance the physical constraint loss and the data matching loss. However, we have seen as we sample more in that region, we reduce the maximum violation of pc-EGP prediction significantly from the order of $10^{-2}$ to $10^{-6}$, whereas the violation of the standard GP prediction increases (Supplementary **Fig A2**). The absolute difference of the pc-EGP predicted highest and the lowest density from the actual simulated mean value are **0.0713** and **0.0012** respectively with tuned weights of $[w_1 = 5, w_2 = 1]$, showing strong improvement compared to the standard GP prediction.

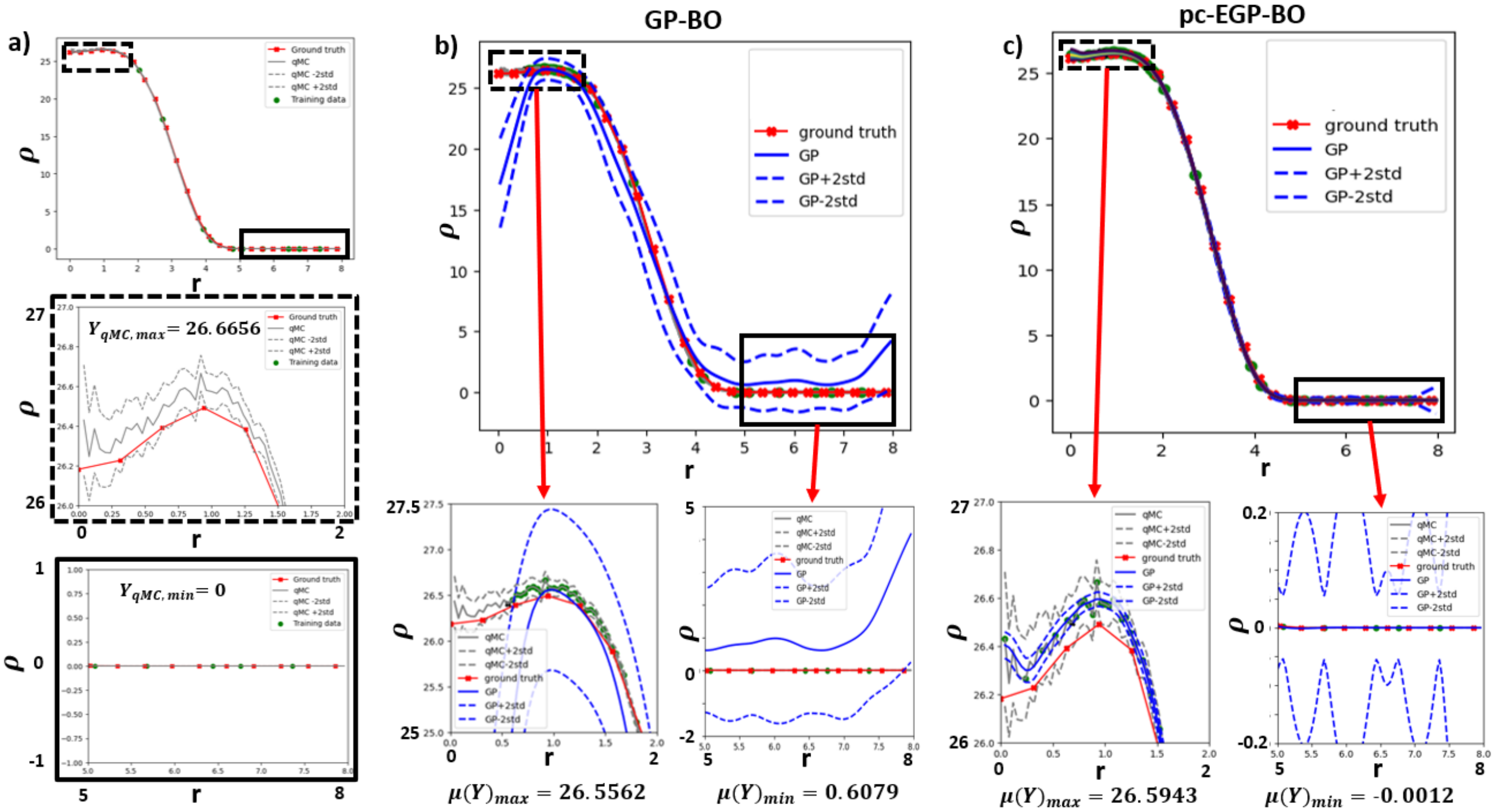

**Fig 4.** Performance of pc-EGP over noisy QMC simulations of the ground state density $\rho$ of $^4$He atoms confined inside a smooth cylindrical nanopore. Fig 4(a) shows the expensive low-sampled ground truth and noisy QMC samples in red and gray lines respectively. The gray dashed lines are the $\pm 2$ standard errors of the QMC simulations. The green dots are the 10 randomly selected samples to start the GP-BO and pc-EGP-BO explorations. Fig 4(b)-4(c) are the prediction (solid blue lines) and $\pm 2$ standard deviation (dashed blue lines) from GP and pc-EGP respectively after 30 autonomous explorations, based on exploiting the high-density region. For pc-EGP, the tuned weights are $[w_1 = 5, w_2 = 1]$.

Having benchmarked the utility of the pc-EGP for a test case of helium in nanopores at zero temperature, we now move to the more challenging task of using pc-EGP-BO to explore and predict over a multi-dimensional parameter space where no ground-truth is available that includes varying both position inside the pore and the chemical potential $\nu$ at finite temperature.

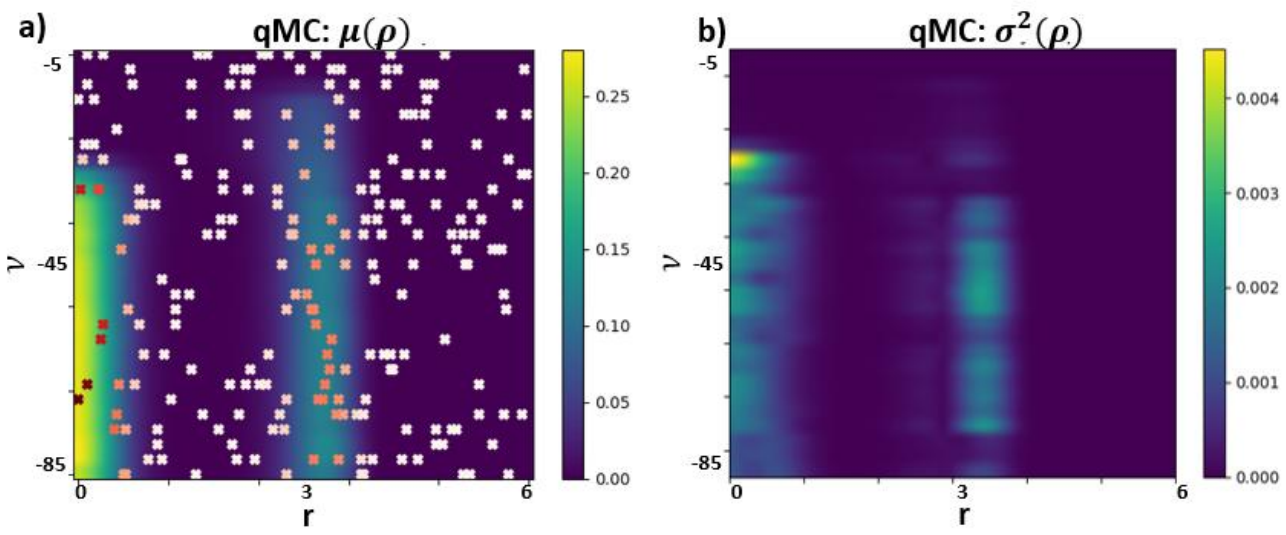

**Fig.5.** QMC simulated ground truth of (a) density $\rho$ and (b) the respective noise/variance at nanopore radius $R = 6.0\, A^o$ over the 2D parameter space of radial position $r$ and chemical potential $\nu$. The dots in fig. 5(a) are the random 220 training samples used for estimation.

**Fig. 5a** shows the expensive QMC simulated ground truth of density $\rho$ at nanopore radius $R = 6.0\,\text{Å}$ over the 2D parameter space of radial position $r$ and chemical potential $\nu$. **Fig. 5b** shows the respective QMC simulation noise (variance), where we can see the higher density region has higher noise. From domain-knowledge, we know that at any value of $(r, \nu), \rho \geq 0$. Out of the 5800 simulated data (each value of $\upsilon$ corresponding to a horizonal slice of **Fig. 5** requires an expensive independent simulation), we randomly choose 220 locations, as marked with crosses in Fig. 5a, for training the GP models for estimation. As can be seen in the figure, the function to be learned, $\rho(r; \upsilon)$ is non-linear, involving multiple regions of maxima and minima. **Fig 6.** shows the performance of the physically meaningful estimation of the proposed model pc-EGP. During the training process of each GP, we considered the Adam optimizer with learning rate of 0.01 for the hyperparameter optimization with total epochs of 75. We consider the Matern Kernel **eq. 5.** From **fig. 6a,** we can see while the standard GP prediction matches best with the ground truth QMC simulation in **Fig. 5a**, it provides unphysical function predictions over as much as 38% of the overall parameter space. Thus, to define a better performance metric, we penalize the mean absolute error with constraint violation by adding a penalty value, $Pen = 0.1$ with the absolute error, $|\rho_{QMC} - \mu(\rho)|$ for any violated estimation $\mu(\rho) < 0$. We choose $Pen = 0.1$ as it is the largest absolute error obtained, to avoid either over- or under-weightage of the penalty. To keep consistency, we fix the penalty value for the rest of the analysis in **fig. 6**. The penalized Mean Square Error, $pMAE$ is $0.044$. **Figs. 6b-6d** are the pc-EGP prediction with increasing weights of physically constrained loss as $w_1 = 2, 5, 12$ respectively. We clearly see a trade-off between better overall prediction (minimizing MAE) and better physically meaningful prediction (minimizing physical constraint violation). We see a significant reduction in infeasible estimate to 11% with $[w_1 = 2, w_2 = 1]$, and 0% with $w_1 = 12$. As the overall prediction error increases, we see the penalized Mean Square Errors, $pMAE$, for these 3 cases are $0.035, 0.033$ and $0.032$ respectively. Thus, to balance both, it appears preferable to choose $w_1 = 2$. However, as preferred by the domain expert objectives, these weights can be controlled which aids the human-in-the loop for alignment in ML-driven autonomous systems.

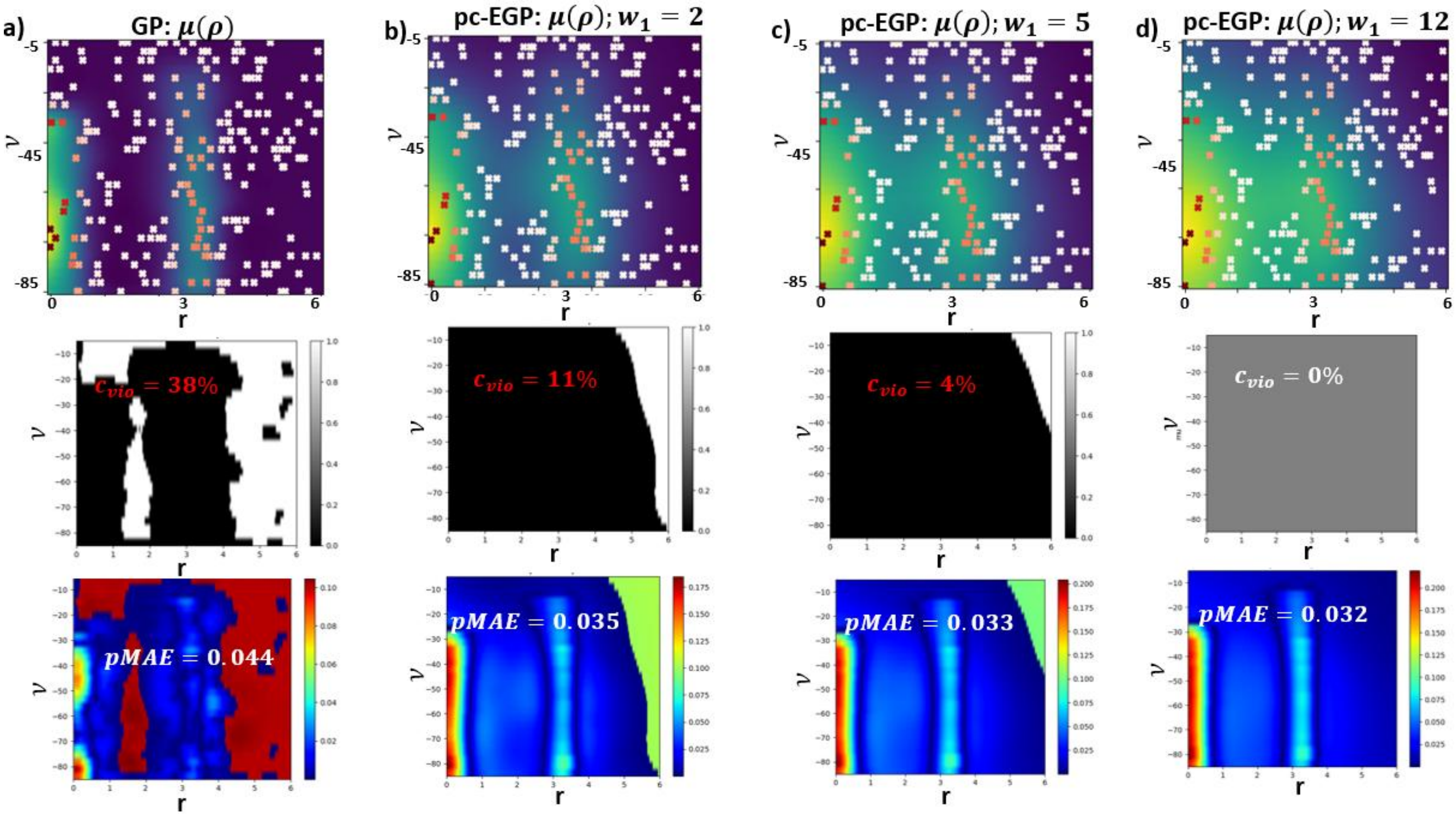

**Fig. 6.** Performance of pc-EGP over noisy QMC simulations of helium confined inside cylindrical nanopores with different filling fractions set by the tunable chemical potential $\upsilon$. Simulation details are included in Ref.[35]. Here, we consider a 2D parameter space of radial position $r$ and chemical potential $\nu$ and the function space of density $\rho$ where the physical constraint on the density is $\rho \geq 0$. Fig 4(a) is the mean density map from GP as trained with 220 samples (referred as $x$) while figs. 4(b)-4(d) are the respective mean density map pc-EGP at different weights of $w_1 = 2, 5, 12$ respectively. For all the cases, $w_2 = 0$. For all the figures 4(a)-4(d), the middle images are the constraint violation map, where the white area is violated $g: \{\mu(\rho) < 0\} = 1$. The total violated percentage is calculated as the ratio between number of predictions violated and the total (5800) number of predictions. The bottom images are the penalized absolute error maps where for any constrained violation a penalty value, $P = 0.1$ is added with the absolute error between QMC simulations and the model predictions. Then after adding penalty, we calculated the mean absolute error, MAE.

## 4. CONCLUSION

To summarize, we have proposed a physically-constrained ensemble Gaussian Process model for improved and physically meaningful learning of computationally expensive quantum systems. The two key developments in the proposed approach, during the model training process, are, designing 1) a pathway for inclusion of physical constraint validation and integration into the loss function (to optimize) and 2) an ensemble approach to train multiple physically constrained GPs simultaneously over different quadrature points defined from heteroskedastic noise of the physical systems. To validate the proposed method, we have demonstrated over different problems such as identifying the one-dimensional Bose-Hubbard model parameter that yields a quantum (non-thermal) phase transition, and a potential physical realization of the one-dimensional Bose-Hubbard model where helium atoms are confined inside an ordered nanoporous glass. The results show improvement in estimation than the pure data-driven

GP method. The proposed approach establishes a robust foundation for autonomous and physically interpretable surrogate modeling of expensive quantum systems and can be easily transferable for meaningful exploration and future exploration decision making over other expensive physical models and autonomous experiments.

**Acknowledgements:**
This research was primarily supported by the National Science Foundation Materials Research Science and Engineering Center program through the UT Knoxville Center for Advanced Materials and Manufacturing (DMR-2309083). A.B. acknowledges the use of facilities and instrumentation at the UT Knoxville Institute for Advanced Materials and Manufacturing (IAMM) and the Shull Wollan Center (SWC) supported in part by the National Science Foundation Materials Research Science and Engineering Center program through the UT Knoxville Center for Advanced Materials and Manufacturing. M. T. acknowledges support by DFG Grant No. RO 2247/16-1, and hospitality from the Center for Advanced Materials and Manufacturing at the University of Tennessee. Computations were performed using resources provided by the Leipzig University Computing Center and University of Tennessee Infrastructure for Scientific Applications and Advanced Computing (ISAAC).

**Code and Data Availability Statement:**
The analysis reported here along with the code is summarized in Notebook for the purpose of tutorial and application to other data and can be found in https://github.com/arpanbiswas52/ papers-code-pcEGP.

Supplementary Materials of the paper titled

# Physically Constrained Ensemble Gaussian Process Modelling for Expensive Quantum Systems with Heteroskedastic Noise

Arpan Biswas[1,2], Sutirtha Paul[2,3], Joseph Agada[4], Matthias Thamm[5], Adrian Del Maestro[2,3]

[1]University of Tennessee-Oak Ridge Innovation Institute, University of Tennessee, Knoxville, TN 37996, USA

[2]Institute for Advanced Materials and Manufacturing, University of Tennessee, Knoxville, Tennessee 37996, USA

[3]Department of Physics and Astronomy, University of Tennessee, Knoxville, TN 37996, USA.

[4]Bredesen Center for Interdisciplinary Research, University of Tennessee, Knoxville, USA, 37996

[5]Institut für Theoretische Physik, Universität Leipzig, 04103 Leipzig, Germany

**Table A1: Bayesian Optimization**

1. **Initialization for BO:** State maximum BO iteration, $M$. Randomly select $m$ samples from the parameter space $\boldsymbol{X}$. Assuming $f$ is the expensive objective function. Set $k = 1$. Evaluate $m$ samples for objective as, $\boldsymbol{Y}(\boldsymbol{X})$. Build training data matrices, $\boldsymbol{D_k} = \{\boldsymbol{X_k}, \boldsymbol{Y_k}\}$.

For $k \leq M$

2. **Surrogate Modelling**: Develop or update GP models, given the training data, as $\Delta(\boldsymbol{D_k})$. **Here, we have integrated our proposed pc-EGP model.**

3. **Posterior Predictions**: Given the surrogate model, compute posterior means and variances for the unexplored locations, $\widehat{\boldsymbol{X_k}}$, as $\boldsymbol{\mu}(\boldsymbol{Y}(\widehat{\boldsymbol{X_k}})|\Delta_j$ and $\boldsymbol{\sigma^2}(\boldsymbol{Y}(\widehat{\boldsymbol{X_k}})|\Delta_j$ respectively. **Here, $\boldsymbol{j}$ GP models of the pc-EGP provide parallel estimations and then we calculate the weighted mean and variances as per eq. 19-20.**

4. **Acquisition function:** Compute and maximize acquisition function, $\max_{x_{best} \in \widehat{X_k}} \Upsilon(f|\Delta)$ to select next best location, $\boldsymbol{x_{best}}$ for evaluations.

5. **Augmentation:** Evaluate $y(\boldsymbol{x_{best}})$. Augment data, $D_{k+1} = [D_k; \{\boldsymbol{x_{best}}, y\}$. Repeat Step 2-5 till convergence.

## Appendix B. Additional Figures

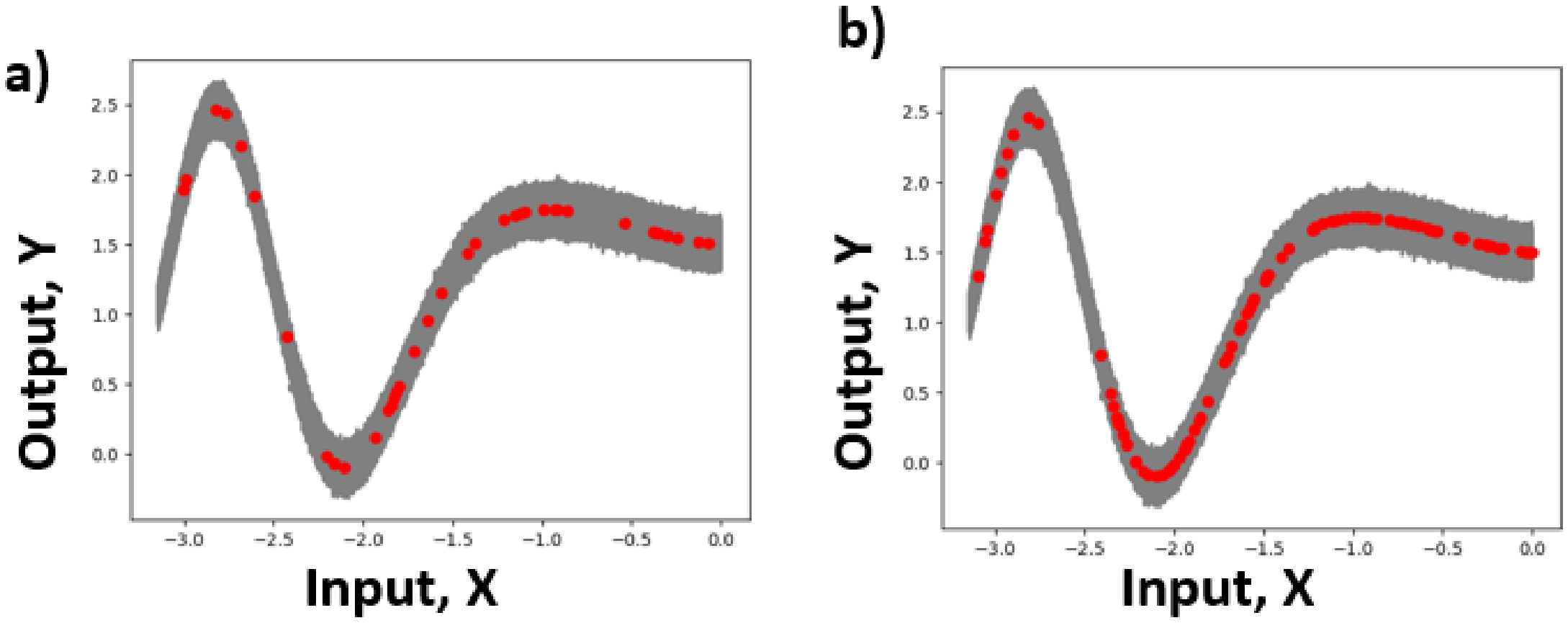

**Figure A1.** Test function with heteroskedastic noise as denoted from the shaded region. The red dots are the sampled training data.

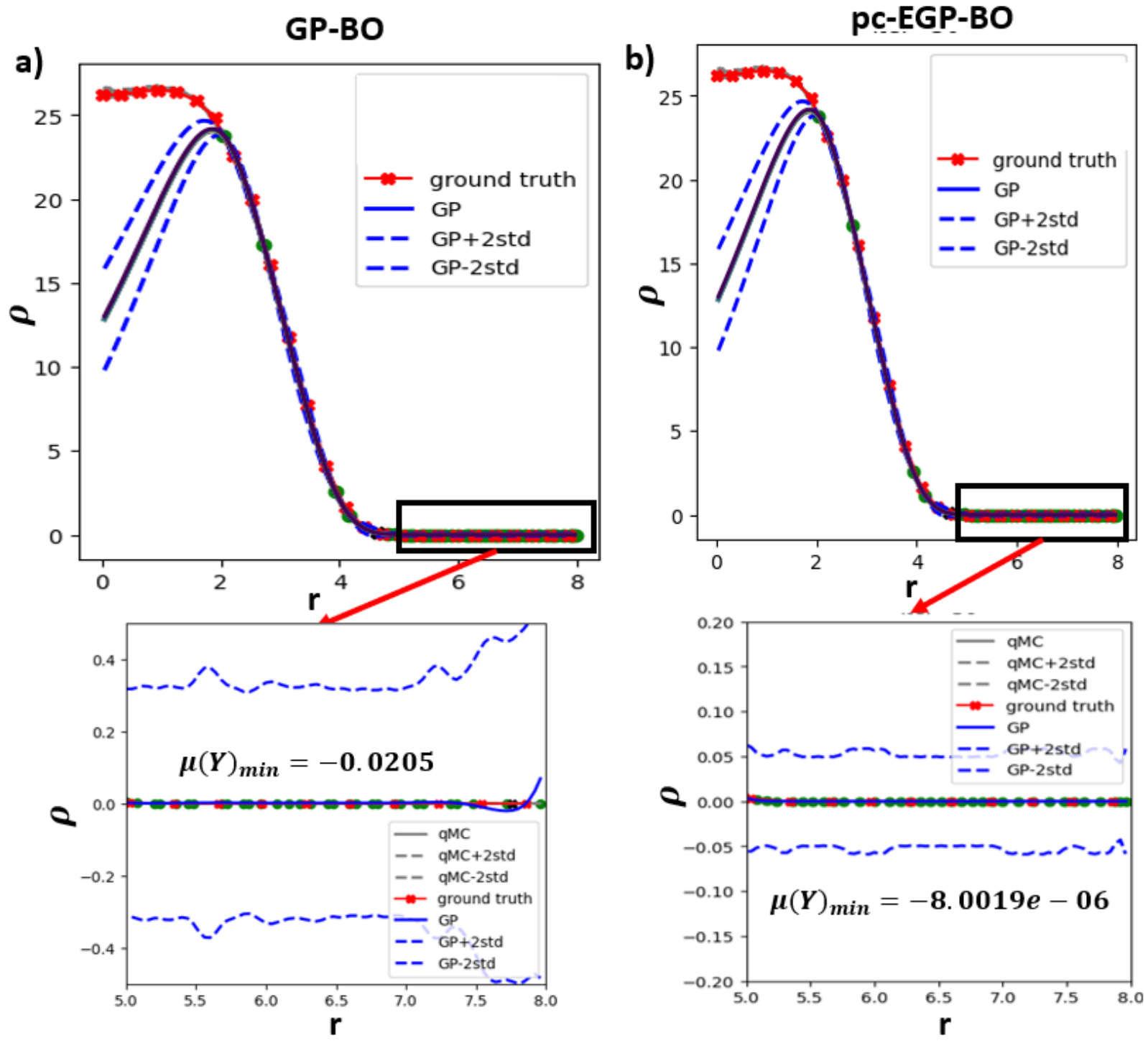

**Figure A2.** Additional Analysis: Performance of pc-EGP over noisy QMC simulations of the ground state density $\rho$ of $^4$He atoms confined inside a smooth cylindrical nanopore. Figs. (a) and (b) are the prediction (solid blue lines) and $\pm 2$ standard deviation (dashed blue lines) from GP and pc-EGP respectively after 30 autonomous explorations, based on exploiting the low-density region. Following the analysis in Fig. 4, we have seen as we sample more in low-density region, the violation of the constraint ($\rho \geq 0$) of pc-EGP prediction significantly reduces from the order of $10^{-2}$ to $10\ {-}^{6}$, whereas the violation of the standard GP prediction is comparatively much higher.